# A GENERAL FORMULATION FOR STANDARDIZATION OF RATES AS A METHOD TO CONTROL CONFOUNDING BY MEASURED AND UNMEASURED DISEASE RISK FACTORS

By Steven D. Mark[1]

*University of Colorado School of Public Health*

Standardization, a common approach for controlling confounding in population-studies or data from disease registries, is defined to be a weighted average of stratum specific rates. Typically, discussions on the construction of a particular standardized rate regard the strata as fixed, and focus on the considerations that affect the specification of weights. Each year the data from the SEER cancer registries are analyzed using a weighting procedure referred to as "direct standardization for age." To evaluate the performance of direct standardization, we define a general class of standardization operators. We regard a particular standardized rate to be the output of an operator and a given data set. Based on the functional form of the operators, we define a subclass of standardization operators that controls for confounding by measured risk factors. Using the fundamental disease probability paradigm for inference, we establish the conclusions that can be drawn from year-to-year contrasts of standardized rates produced by these operators in the presence of unmeasured cancer risk factors. These conclusions take the form of falsifying specific assumptions about the conditional probabilities of disease given all the risk factors (both measured and unmeasured), and the conditional probabilities of the unmeasured risk factors given the measured risk factors. We show the one-to-one correspondence between these falsifications and the inferences made from the contrasts of directly standardized rates reported each year in the *Annual Report to the Nation on the Status of Cancer*. We further show that the "direct standardization for age" procedure is not a member of the class of unconfounded standardization operators. Consequently, it can, and usually will, introduce confounding when confounding is not present in the data. We propose a particular standardization operator, the SCC operator, that is in the class of unconfounded operators. We contrast the mathematical properties of the SCC and the SEER operator (SCA), and present an analysis of SEER cancer registry data that demonstrates the consequences of these differences. We further prove that the SCC operator is a projection operator. We discuss how this property can enable the SCC operator to be developed as a method for comparing nested conditional expectations in the same manner as is currently done with regression methods that control for confounding.

Received January 2008; revised January 2008.

[1]Supported by the University of Colorado Health Sciences Center at Denver.

**1. Introduction.** Each year the NCI's Surveillance, Epidemiology, and End Results (SEER) program compiles data (henceforth called SEER data) on cancer incidence and mortality from (currently) 17 population-based cancer registries in the United States [Howe et al. (2006)]. Since 1998 the National Cancer Institute, the American Cancer Society, the Centers for Disease Control, and the North American Association of Central Cancer Registries have analyzed the SEER data to produce an *Annual Report to the Nation on the Status of Cancer in the United States* (subsequently referred to as the *Annual Reports*). These reports contain estimates of the overall annual cancer incidence and mortality, as well as incidence/mortality by cancer site, and incidence/mortality within population subgroups defined by gender, race, ethnicity, and geographic location of the cancer registry. Some of the stated goals of these reports are to: (1) report on the cancer burden as it relates to cancer incidence and mortality and patient survival; (2) identify unusual changes and differences in the patterns of occurrence of specific forms of cancer in population subgroups defined by geographic, demographic, and social characteristics; (3) describe temporal changes in cancer incidence, mortality, extent of disease at diagnosis (stage), therapy, and patient survival that might impact cancer prevention and control strategies; (4) monitor the occurrence of possible iatrogenic cancers; and (5) attribute changes in cancer rates to temporal changes in diagnostic criteria, screening, preventive measures, cancer treatments, or environmental exposures [SEER (2005a), Ward et al. (2006)]. In addition to the goals common to all of the *Annual Reports*, each report has a special sub-focus. Since 2001 these reports have stated conclusions regarding: (1) absolute population rates and changes in cancer rates [Howe et al. (2001), Edwards et al. (2002), Weir et al. (2003), Jemal et al. (2004), Edwards et al. (2005), Howe et al. (2006)]; (2) the impact of screening and treatment on specific cancers [Howe et al. (2001)]; (3) differences in cancer rates by gender, race, ethnicity, and geographic location [Howe et al. (2001), Weir et al. (2003), Jemal et al. (2004), Edwards et al. (2005)]; (4) causes of the difference in rates within the subgroups listed in (3) [Jemal et al. (2004), Edwards et al. (2005), Howe et al. (2006)]; and (5) the future public policies and expenditures that should be undertaken to increase cancer prevention and improve access to medical care [Weir et al. (2003), Edwards et al. (2005), Howe et al. (2006)].

In order to make meaningful statements about the year-to-year changes in cancer incidence/mortality as a function of one set of characteristics, it is necessary to control for differences in the frequency of cancer risk factors that are not in the set of interest. We refer to any statistical procedure that attempts to separate the effect on cancer rates of one set of measured covariates from another set as procedures that control for confounding. The common methods of controlling for confounding are as follows: (1) multivariate regression; (2) stratification; and (3) standardization. Standardization is virtually always the method of choice when inference is made from population-studies, or data from disease registries. It is the procedure used in the *Annual Reports* [Klein and Schoenborn (2001), Ries and Kosary (2005)].

The particular standardization method used to analyze SEER data is designed to control for year-to-year differences in age distributions. We refer to this method as **Standardization Controlling for Age (SCA)**. In this paper we present a new procedure that allows researchers to control for any set of measured covariates. We refer to this procedure as **Standardization Controlling for Covariates (SCC)**.

The paper is organized as follows. In Section 1 we define nomenclature for a completely general data structure, and describe the SEER data in terms of this nomenclature. In Section 2 we give formulae for the usual representations of standardized rates as weighted averages of a given set of stratum specific rates. We detail the rationale for the specific choice of weights used in SCA standardization. We then define SCA and SCC standardized rates as the output of SCA and SCC operators. These operators are functionals of the empirical distribution of a given set of data, and a user-defined weighting distribution. Using the operator formulation, we define a general class of all standardization operators. In Section 3 we formalize our previous discussion of the goals of standardization. We define criteria that specify when contrasts of crude-cancer rates are "not confounded." We extend these criteria to define the subclass of standardization operators that produce contrasts of standardized rates that are not confounded. The SCC operator falls within this subclass; the SCA operator does not. We show that if one begins with crude rate differences that are not confounded, the SCA operator introduces confounding. We discuss how the differences in properties of the SCA and SCC operators relate to the differences in the functionals. In Section 4 we present analyses of the SEER 13 data that demonstrate the properties described in Section 3.

Up until Section 5 we discuss confounding in terms of measured risk factors. In Section 5 we provide a formal framework for examining what inferences can be made from the standardized rate differences produced by the SCC operator in the presence of unmeasured risk factors. Such inferences require assumptions that can be neither completely falsified nor confirmed by examination of the observed distribution functions. We show that nonzero between-year differences in standardized rates allow one to reject certain assumptions about unmeasured risk factors, and that violations of these assumptions correspond directly to the inferences made by SEER investigators in the *Annual Reports* [Ward et al. (2006)].

In Section 6 we change focus from between-year inferences to within-year inferences. We define "nested" standardized rates, derive the properties of nested rates produced by the SCA and SCC operators, and discuss implications for within-year model building of standardized rates.

In the discussion section we summarize our results, suggest how the standardization operators we propose can be used for nonparametric, semiparametric, or parametric estimation of conditional means, and discuss the direction of our current work on developing software that will implement the operators we describe.

**2. Data structure, crude-cancer rates and finest-crude-cancer rates.** Let $(D^y, Z^y)$ be any vector of real valued random variables, and $P^y(D, Z)$ any set of probability distributions defined on the support of $(D^y, Z^y)$; $y \in \mathcal{Y}$; $\mathcal{Y} \equiv \{1, 2, \ldots, n\}$; $n$ finite. Since the support of $P^y(D, Z)$ does not vary with $y$, we frequently drop the superscripts for random variables and write $(D, Z)$. In SEER data $D$ is a vector of indicator variables denoting the presence or absence of a specific form of cancer type; $Z$ is the set of all other measured covariates; $P^y(D, Z)$ is the empirical distribution of $(D^y, Z^y)$ for year $y$; $\mathcal{Y}$ is the set of years for which we have data on $(D, Z)$. Thus, the SEER data for year $y$ consists of an observation of $P^y(D, Z)$. Since in the *Annual Reports* an individual is classified as either having or not having a specific cancer (or a cancer in a defined set), and the joint distribution of cancers is not of interest, we will regard $D$ to be a binary random variable: $D = 1$ when an individual is in the set of cancers of interest, $D = 0$ otherwise. We assume that our interest is in the distribution $P^y(D, E)$, where $E$ is the (possibly improper) subset of $Z$ which investigators believe contain all the "measured cancer risk factors." Without loss of generality we adhere to the structure of the SEER data, and regard the support of $(D, E)$ as discrete. We assume that in SEER the measured cancer risk factors, $E$, consist entirely of information about an individual's age, gender, race, ethnicity, and catchment area of cancer registry (henceforth called place):

(1) $$E = (\text{age, gender, race, ethnicity, place}).$$

These are, in fact, the only covariates required to produce the standardized rates given in the *Annual Reports*.

Let $E = (E_1, E_2)$ be any factorization of $E$ such that $E_1 \cap E_2 = \varnothing$; $E_1 \subseteq E$. We use notation such as $P^y(D|E)$, $P^y(D|E_1)$, $P^y(E_2|E_1)$ to denote the conditional distributions of $P^y(D, E)$. For any $E^\dagger \subseteq E$, we refer to $P^y(D|E^\dagger)$ as the **crude-cancer rate** for $E^\dagger$. When $E^\dagger = E$, we refer to $P^y(D|E)$ as the **finest-crude-cancer rate**.

Any crude-cancer rate is related to the finest-crude-cancer rate by the integral given in (2).

(2) $$P^y(D|E_1) = \int_{\mathcal{E}_2} P^y(D|E_1, E_2) \, dP^y(E_2|E_1).$$

The region of integration in (2) is $\mathcal{E}_2$, the support of $E_2$. Throughout the paper calligraphic letters indicate the support of random variables. When as in the SEER data, $(D^y, E^y)$ have discrete support, we can express the right-hand side of (2) as a sum of the product of discrete conditional probabilities,

$$P^y(D|E_1) = \sum_{\mathcal{E}_2} P^y(D|E_1, E_2) \times P^y(E_2|E_1).$$

The crude-cancer rate on the left-hand side is the frequency of disease in subjects with a give value of $E_1$.

2.1. *General definition and formulae for standardized rates.* We define $s_y^*[D|E_1]$ to be a **standardized cancer rate given** $E_1 = e_1^*$ if it can be expressed in the form of the integral given in (3),

$$(3) \qquad s_y^*[D|e_1^*] \equiv \int_{\mathcal{E}_2^\dagger} P^y(D|e_1^*, e_2^\dagger) \, dP^*(e_2^\dagger).$$

Here $E_2^\dagger \subseteq E$, $e_2^\dagger \in E_2^\dagger$, and $P^*(E_2^\dagger)$ is any user-defined measure that has the same support as $E_2^\dagger$ and is consistent with a probability measure. Under the restriction that $(D, E)$ has discrete support, we can write (3) as

$$(4) \qquad s_y^*[D|e_1^*] \equiv \sum_{e_2^\dagger \in \mathcal{E}_2^\dagger} P^y(D|e_1^*, E_2^\dagger = e_2^\dagger) \, dP^*(e_2^\dagger).$$

Equation (4) is the weighted sum of stratum specific weights: $E_2^\dagger$ define the strata; $dP^*(e_2^\dagger)$ is the weight for stratum $e_2^\dagger$; $P^y(D|e_1^*, E_2^\dagger = e_2^\dagger)$ is the crude-cancer rate within stratum $E_1 = e_1^*, E_2^\dagger = e_2^\dagger$. Equation (4) is equivalent to the usual algebraic definition of a standardized rate [Rothman (1986)].

Typically, discussions about standardization assume the strata are fixed and focus on the choice of weights [Rothman (1986)]. The general advice is that the choice of weights should depend upon the interpretation one desires to ascribe to the standardized rates [Rothman (1986)]. The weights used in the SEER implementation of the SCA method are the age-frequency of the US population in year 2000 [Klein and Schoenborn (2001), Ries, Eisner, and Kosary (2005), Ward et al. (2006)]. This is referred to as direct standardization [Klein and Schoenborn (2001), Rothman (1986)]. In particular, the SCA procedure of SEER is described as: "Age adjustment, using the direct method, is the application of observed age-specific rates to a standard age distribution to eliminate differences in crude rates in populations of interest that result from differences in the populations' age distribution [Klein and Schoenborn (2001)]." The justification for direct standardization of the cancer rates in year $y$ is that the standardized rate for year $y$ will represent the cancer rates that would have been observed in year $y$ had the age distribution in year $y$ been identical to the age distribution in year 2000 [Klein and Schoenborn (2001), Rothman (1986), Anderson and Rosenberg (1998)]. The advantage of expressing standardized rates in terms of "what would have been seen in some year $y$" is that such weighting produces standardized rates which preserve the magnitude of the crude-cancer rates. Since the magnitude of the year-to-year differences in cancer rates are of importance to the inferences made in the *Annual Reports*, it is desirable to choose a standardization procedure that preserves these values.

2.2. *Defining the SCA and SCC operators.* To contrast the properties of SCA and SCC standardized rates, it is best to regard them as the output of SCA and SCC operators. We define a standardization operator, $S_y^*[D|E_1]$, to be any functional of $P^y(D, E)$, $E_1$, $E_2^\dagger$ and a user-defined probability distribution, $P^*(E_2^\dagger)$, that can be expressed by the integral in (5):

$$(5) \qquad S_y^*[D|E_1] \equiv \int_{\mathcal{E}_2^\dagger} P^y(D|E_1, E_2^\dagger) \, dP^*(E_2^\dagger).$$

For a particular $E_1 = e^*$, the standardized rate is denoted by the left-hand side of (3).

Let $P^*(E)$ be a user-defined probability distribution with the same support as $E$. Let $A$ be any random variable in $E$, and $P^*(A)$ the marginal probability of $A$ from distribution $P^*(E)$. We denote the points of support of $A$ as $(a_1, a_2, \ldots, a_N)$. Using "\" as the set difference operator, we define $E^a = E \setminus A$, $E_1^a = E_1 \setminus A$; $E_2^a = E_2 \setminus A$.

We define the **SCA operator**, $S_y^{ca}[D|E_1^a]$, to be

$$(6) \qquad S_y^{ca}[D|E_1^a] \equiv \int_A P^y(D|E_1^a, A) \, dP^*(A).$$

$s_y^{ca}[D|e_1^*]$ denotes the SCA standardized rate given $E_1^a = e_1^*$.

In the *Annual Reports* standardized rates are produced using the SCA operator: $A$ is age, and the support points are the five year intervals into which age is categorized. The weighting distribution, $P^*(E)$, is the covariate distribution in year 2000, $P^{2000}(E)$. Thus, $P^*(A)$ is the age frequency in year 2000.

For example, let $D$ be colon cancer, $E_1^a = $ (gender), and $E_2^a = $ (race, ethnicity, place). The SEER SCA estimate of the standardized rate of colon cancer conditional on gender being male is

$$s_y^{ca}[\text{colon cancer} \mid \text{male}] = \sum_{j=1}^{N} P^y(\text{colon cancer}|\text{male, age} = a_j)$$
$$\times P^{2000}(\text{age} = a_j).$$

Here $P^y$ (colon cancer | male, age $= a_j$) is the frequency of colon cancer in year $y$ for males in age category $a_j$, and $P^{2000}$(age $= a_j$) is the frequency of age group $a_j$ in year 2000.

We define the **SCC operator**, $S_y^{cc}[D|E_1]$, to be the functional of $P^y(D, E)$, $P^*(E)$, and $E_1$, given by the integral in (7):

$$(7) \qquad S_y^{cc}[D|E_1] \equiv \int_{\mathcal{E}_2} P^y(D|E_1, E_2) \, dP^*(E_2|E_1).$$

Using the same factorization of $E$, and the weighting distribution specified by (7), the SCC estimate of the standardized colon cancer rate conditional on gender equals male is

$$\begin{aligned} s_y^{cc}&[\text{colon cancer} \mid \text{male}] \\ &= \sum_{\mathcal{E}_2} P^y(\text{colon cancer} \mid \text{male, age, race, ethnicity, place}) \\ &\quad \times dP^{2000}(\text{age, race, ethnicity, place} \mid \text{male}). \end{aligned}$$

Here $P^y$ (colon cancer | male, age, race, ethnicity, place) is the frequency of colon cancer in year $y$ for males within strata defined by age, race, ethnicity, and place; $dP^{2000}$(age, race, ethnicity, place | male) is the frequency among males for a given (age, race, ethnicity, place).

For completeness we note that we have explicitly presented the SCA and SCC operators in terms of the random variables $(D, E)$ and the empirical distributions $P^y(D, E)$. If we restrict the data set to some $D^\ddagger \subset D$, and/or $E^\ddagger \subset E$, the operators are defined in terms of the empirical distribution $P^y(D^\ddagger, E^\ddagger)$. In practice, given the strong correlation of cancer type with age, such analyses are common. In Section 4 we present standardized rates for colon cancer and breast cancer. These rates were made from data limited to subjects 40 years of age or older. Similarly, when Ward et al. report standardized rates for childhood cancers, they restrict subjects to those age 19 or less.

**3. Contrasting the properties of the SCC and SCA operators.** Inspection of the formulas for the SCA (6) and SCC (7) operators reveals two important differences. In the SCA operator the crude-cancer rate varies as a function of $E_1$, but the weight does not. In the SCC operator the crude rate is always the finest-crude-cancer rate, and the weight depends on $E_1$. In this section we examine the consequences of these differences on the properties of the standardized rates. We begin by formalizing the goals of standardization discussed at the end of Section 2.1.

3.1. *Standardization operators and the control of confounding by measured risk factors.* The *Annual Reports* compare year-to-year differences in cancer rates conditional on some subset $E_1$ of $E$ (1). The need for standardization arises because of the concern that differences in the crude-cancer rates may reflect year-to-year differences in the distribution of $E_2$. From (2), we see that the distribution of $E_2$ that affects the crude-cancer rate is $P^y(E_2|E_1)$.

If for years $y^\dagger$ and $y^{\dagger\dagger}$,

$$(8) \qquad P^{y^\dagger}(E_2|E_1) = P^{y^{\dagger\dagger}}(E_2|E_1),$$

we say **there is no $E_2$ confounding of the $E_1$ crude rate differences**,

$$(9) \qquad P^{y^\dagger}(D|E_1) - P^{y^{\dagger\dagger}}(D|E_1).$$

When (8) is true, contrasts of the crude rates provide the best rates of the year-to-year differences. From (8) we know that the differences in the crude-cancer rates cannot be due to differences in the $E_2$ distribution; trivially, the crude rate differences achieve the desired goal of having the standardized contrasts preserve the observed magnitude of the differences in crude rates.

Assume now that (8) is true for all $y \in \mathcal{Y}$, and that the weighting distribution used is $P^{2000}(E)$. By inspection of (7), we see that for all $y$, and any factorization of $E$, standardized rates produced by the SCC operator equal the crude-cancer rates. Thus, if there is no $E_2$ confounding of the $E_1$ crude rate differences, and one uses the SCC operator, contrasts of the SCC standardized rates are contrasts of the unconfounded crude rates.

To see that this is not the case for the SCA operator, we re-express (6) in terms of the finest-crude-cancer rates:

$$(10) \quad S_y^{ca}[D|E_1^a] = \int_A \left\{ \int_{\mathcal{E}_2} P^y(D|E_1^a, E_2^a, a) \, dP^y(E_2^a|E_1^a, a) \right\} dP^*(A=a).$$

Suppose in (10) that $y = 2000$. Even were (8) true, and $P^*(A) = P^{2000}(A)$, the SCA operator does not return the crude-cancer rate. Thus, contrary to the stated justification for the choice of weights, the SCA standardized rates in year 2000 conditional on $E_1^a$ do not equal the observed cancer rates in year 2000, even though the weights used are the age distribution of year 2000. This will be graphically demonstrated in Figure 1.

Consistent with our definition of no confounding of crude rate differences (8), we define standardized rate differences,

$$(11) \qquad s_{y^\dagger}^*[D|e_1^*] - s_{y^{\dagger\dagger}}^*[D|e_1^*],$$

to be unconfounded by $E_2$, if the standardized rates are produced by a standardization operator, $S_y^*[D|E_1]$, that can be expressed as an integral of the finest-crude-cancer rates with respect to a measure that depends only on the factorization of $E$ [see (A.1) for formal definition]. We refer to such operators as **standardization operators with no $E_2$ confounding (SONC operators)**. The SCC operator is one such operator. In fact, if we chose $P^*(E) = P^{2000}(E)$, it is the unique standardization operator that produces the standardized cancer rates that "would have been seen in year $y$ had the covariate distribution in $y$ been identical to the covariate distribution in year 2000." Note that SONC operators with no $E_2$ confounding produce standardized rate differences that are not confounded by $E_2$ regardless of whether (8) is true.

It is clear from (10) that the SCA operator does not, in general, produce standardized rate differences that are unconfounded. In fact, for a given factorization of $E$ and for specific $y^\dagger$, $y^{\dagger\dagger} \in \mathcal{Y}$, the SCA operator produces unconfounded standardized rate differences if and only if

$$(12) \qquad P^{y^\dagger}(E_2^a | E_1^a, a) = P^{y^{\dagger\dagger}}(E_2^a | E_1^a, a).$$

Since (8) does not imply (12), the SCA operator can produce standardized rate differences that are confounded by $E_2$ even when (8) is true and the crude rate differences are not confounded.

For SCA to produce unconfounded standardized rate differences for all factorizations of $E$ requires (13):

$$(13) \qquad P^{y^\dagger}(E^a | a) = P^{y^{\dagger\dagger}}(E^a | a).$$

When (13) is true for all possible combination of years, $(y^\dagger, y^{\dagger\dagger})$, then conditional on age, the distribution of the other risk factors are identical for all years. Thus, for the SCA operator, which "standardizes only for age," to produce unconfounded standardized rate differences requires that the $P^y(E)$ distributions are identical except possibly for the marginal distribution of age. The equality in (13) does not exist for the analysis of the SEER data we present in Section 4.

If the $P^*(E)$ and $P^y(E)$ distributions are such that

$$(14) \qquad P^*(E^a | A) = P^y(E^a | A) \quad \text{and} \quad P^y(E^a | A) = P^y(E^a),$$

then the rates produced by the SCA and SCC operator are identical. Thus, if (8) is true, the SEER SCA operator always returns the crude-cancer rates iff the $E^a$ distributions are identical for all years and age is independent of $E^a$.

It is instructive to consider the properties of the SCA and SCC operators with regard to confounding by measured risk factors in terms of the familiar regression model approach to control for such confounding. In this context, we regard the $P^y(D, E)$ as random samples from larger populations. For concreteness, we conceptualize that the $P^y(D|E)$ have log-linear Poisson distributions. Given that age is the primary determinant of cancer rates, we propose Poisson models stratified on age category, and obtain estimates of $P^y(D|E)$ from weighted sums of the predicted probabilities in each age stratum. Our goal is to find the most parsimonious Poisson models from which to make inferences regarding the effect of sex and race on within- and between-year cancer rates. We begin with the within-age stratum models saturated in (sex, race, ethnicity, place). The predicted probabilities from these saturated models are identical to the probabilities given by the finest-crude-cancer rates, $P^y(D|E)$. We would make inferences from the more parsimonious models saturated only in (age, sex, race) if: (1) the regression coefficients for all covariates that are a function of either ethnicity or place were equal to zero; or (2) within each age stratum, (sex, race) were (in the data) statistically independent of (ethnicity, place). These criteria correspond to the usual rubric that covariates $E_2$ are not confounders of the effect of $E_1$ provided either $E_2$ are not risk factors for disease, or $E_2$ are uncorrelated with $E_1$.

The conditions required for a standardization operator to be in the class of SONC operators are related to, but more stringent than, those given above. SONC operators are integrals of the finest-crude-cancer rates with respect to a measure that does not depend on year (A.1). If the first regression criteria for non-confounding were true, then the integrand in the SCA operator (6) would be equivalent to the finest-crude-cancer rates. However, the second criterion specifies only that $P^y(E_2^a|E_1^a, a) = P^y(E_2^a|a)$, not that $P^y(E_2^a|a)$ is invariant to year. The latter is clearly necessary for SCA to be a SONC operator (10).

**4. Using the SCA and SCC operators to analyze SEER data.** In this section we present results from our analyses of the SEER data from 13 registries, years 1992–2003 [SEER 13 Regs Limited_Use (2005b); henceforth called SEER 13]. We consider the subset of SEER 13 where age is greater than or equal to 40, race is limited to black or white, and ethnicity is limited to either Hispanic or non-Hispanic. Because of the restriction we place on race, we exclude Alaska and consider only 12 of the 13 cancer registries. In our analyses we limit the covariates to the $E$ defined in (1). The intent of this section is to demonstrate the existence of differences in the standardized rates produced by the SCA and SCC operators, particularly those differences discussed in Section 3. We make no comments about the statistical significance of these findings and provide no formal estimates of trends (see Discussion in Section 7). All rates given are per 100,000 persons.

Figure 1 is a graph of the crude and standardized (SCA, and SCC) race-and-gender specific colon cancer incidence for each year from 1992 to 2003. For the SCA and SCC operators all rates are produced with $P^*(E) = P^{2000}(E)$.

The SCA (dashed line) and SCC (solid line) rates differ for all groups and all years. Thus, (14) is false. For the SCA operator to produce rate differences that are not confounded, (12) must be true for this factorization of $E$. In SEER 13 (12) is false: there exists variation in the year-to-year $P^y$(Hispanic, place|age, gender, race = white). During the time period 1992 to 2003 the frequency of Hispanic ethnicity increased in every place, for both genders, and (with rare exception) for every age group (data not shown). Note, however, that in Figure 1 the slope of each segment of the SCA and SCC plots for white males, and both white and black females, are virtually identical. This indicates that though SCA rate differences may be confounded (11), inferences about the existence of trends may be identical. The slope of the SCA curve is determined by between-year differences in the value of the integral of the finest-crude-cancer rates with respect to the measure $P^y(E_2^a|E_1^a, a)$ (10); confounding of SCA differences requires only year-to-year variation of $P^y(E_2^a|E_1^a, a)$.

The principal motivation for using the weights specified by direct standardization is the desire to produce standardized rates that reflect the true

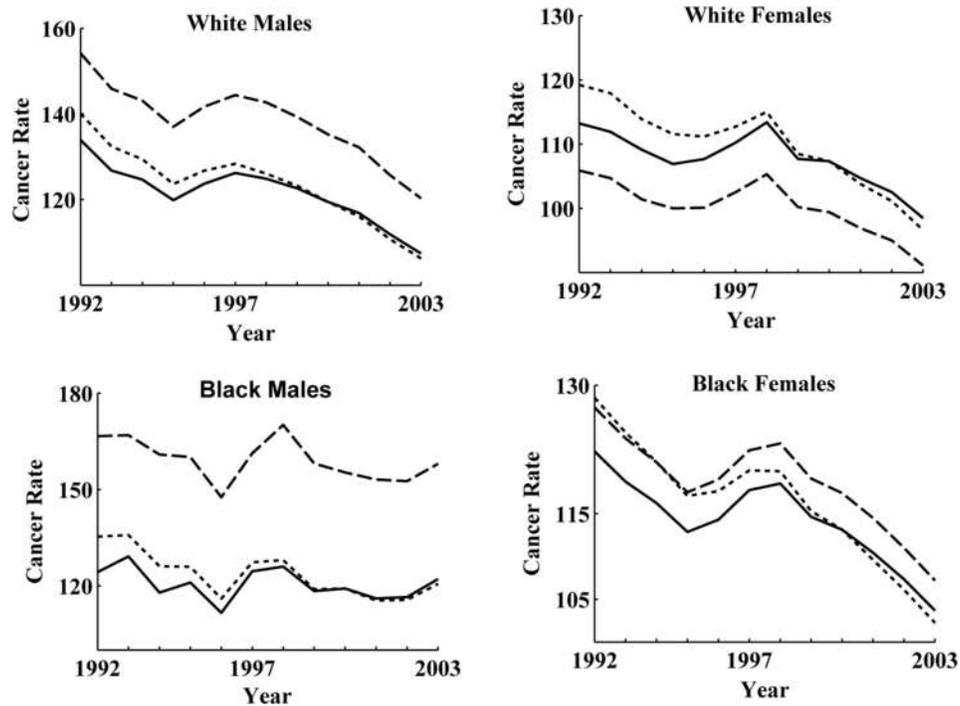

FIG. 1. *Comparing sex and age specific crude-cancer rates with SCA and SCC standardized rates for the years 1992–2003. The dotted line is the crude colon cancer rate. The dashed line is the SCA standardized colon cancer rate. The solid line is the SCC standardized colon cancer rate.*

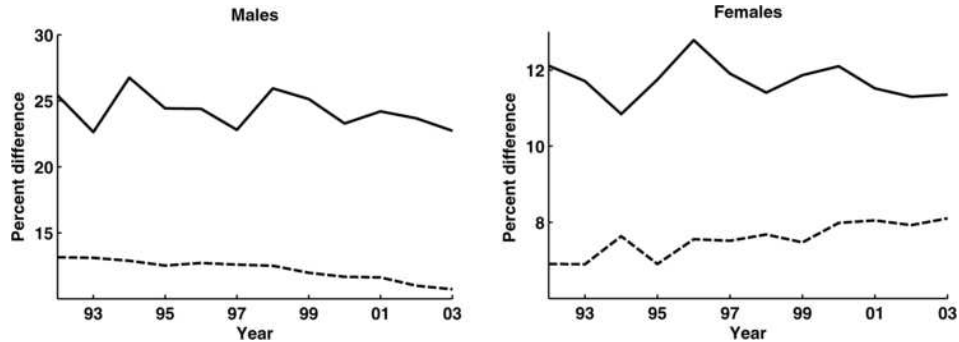

FIG. 2. *The absolute value of the percent difference between the SCA standardized colon cancer rate and the crude colon cancer rate by sex and race for the years 1992–2003. The plotted values for each year were produced by the following formula:*

$$\frac{|SCA\ standardized\ colon\ cancer\ rate - Crude\ colon\ cancer\ rate|}{SCA\ standardized\ colon\ cancer\ rate} \times 100.$$

*The dashed lines are the absolute value of the percent differences for whites. The solid lines are the absolute value of the percent difference for blacks.*

absolute values of the crude-cancer rates in the $E_1$ group. In Figure 1 we see that standardized rates from the SCC operator more closely track the crude-cancer rates than those produced by the SCA operator. In fact, as indicated earlier, the SCA standardized rate in the year 2000 does not equal the crude-cancer rate that would have been seen if the age distribution were identical to that of the year 2000.

Figure 2 is a graph of the magnitude (the absolute value) of the percent difference in the SCA and crude-cancer rates. For both males and females the magnitude of these differences is greatest for blacks (solid line). This phenomenon is due to the fact that the $P^{2000}(\text{age})$ distribution is much closer to the age distribution for whites than for blacks. In addition, the year-to-year variation in the percent deviation appears to be greater for blacks. Thus, graphically, blacks always appear to have larger year-to-year changes in cancer incidence than do whites. These differences are more prominent when cancer rates are compared within groups defined by ethnicity (data not shown). Empirically we find that the lower the population frequency of a group, the greater the deviation of SCA standardized rates from the actual crude-cancer rates.

One previously unmentioned limitation of the SEER SCA operator is that it cannot produce age-specific cancer rates that control for differences in the distribution of other risk factors. Since age is by far the largest risk factor for cancer, comparing within-age-strata rates may reveal trends that are otherwise not visible. Figure 3 is a graph of the SCC standardized breast cancer rates for white females for each year from 1992 through 2003. This graph indicates an overall increase in breast cancer rates from 1992 to 1998, and a decrease from 1998 to 2003. Figure 4 contains a plot of the SCC breast cancer rates for white females in the years 1992 (dotted line), 1997 (solid line), and 2003 (dashed line), within each of the five-year age categories 40 years old or greater. The shape of the graphs of standardized rates are similar for all three years. Consistent with Figure 3, we see that the lowest rates are for year 2003, and the highest for 1997. What cannot be discerned from Figure 3 is that the differences in cancer rates for the years 2003 and 1997 are greatest for females older than 60; and that the differences between 2003 and 1992 rates are almost entirely due to rate differences in females over 60. The information in Figure 4 suggests that when considering possible causes of the calendar trends shown in Figure 3, one should focus on changes that were more prominent in females age 60 and older. Given that the usual approach in epidemiology is to make inferences about cancer rates within age groups, and not from marginal rates obtained by integrating out age, we anticipate that age-specific standardized rates will provide important additional information about cancer trends and between subgroup differences in cancer rates.

Note that if one were to employ a "stratification strategy" and use the SCA operator to calculate separate standardized rates for each age group, the age standardized rates produced would in fact be the crude breast cancer rates, $P^y$(breast cancer | white, female, age $= a_j$).

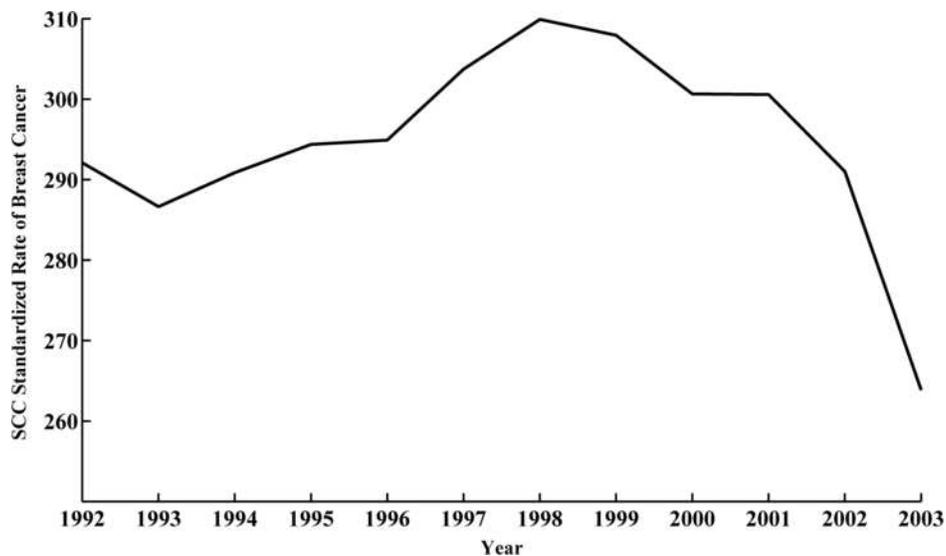

Fig. 3. *The SCC standardized breast cancer rates for white females age forty and older for the years 1992–2003.*

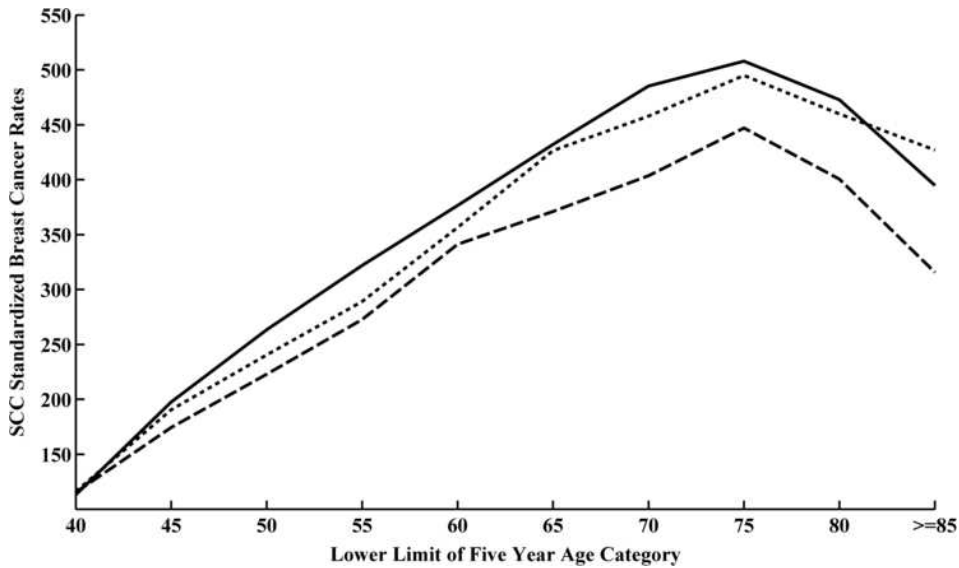

Fig. 4. *The SCC standardized breast cancer rates by five-year interval for white females age forty and older. The dotted line is the SCC standardized breast cancer rates for 1992. The solid line is the SCC standardized breast cancer rates for 1997. The dashed line is the SCC standardized breast cancer rates for 2003.*

**5. Making inferences from observed rate differences of SCC standardized rates.** In Section 3 we established that the SCC operator is in the class of operators in which differences in the $P^y(E_2|E_1)$ do not affect the standardized rates produced by those operators. Thus, if the SCC standardized rates for year $y^\dagger$ and $y^{\dagger\dagger}$ differ, we can conclude that the finest-crude-cancer rates differ for those years. However, despite the nomenclature, we do not know whether for fixed $E_2$ the finest-crude-cancer rates differ as a function of $E_1$, for fixed $E_1$ they differ as a function of $E_2$, or whether the differences in finest-crude-cancer rates depend on both $E_1$ and $E_2$.

To make the inferences of interest to the SEER investigators [Ward et al. (2006)], we require that we consider disease rates as a function of both measured and unmeasured risk factors. To incorporate the effect of unmeasured risk factors on inference, we use the fundamental disease probability (FDP) paradigm for inference proposed by Mark (2004, 2005, 2006, 2008). The results we present in this section depend only on the definitions presented in this section, and require no knowledge of, or results from, any of the other material contained in Mark (2004, 2005, 2006, 2008).

We define the fundamental disease probability for year $y$ to be the probability of disease conditional on all risk factors. We denote this by $P^y(D|E,U)$. Here $E$ are the measured risk factors; $U$ is a set of unmeasured risk factors that, along with $E$, completely determine the probability of disease. The relationship between the FDP and the finest-crude-cancer rates, $P^y(D|E)$, is

$$(15) \qquad P^y(D|E) = \int_U P^y(D|E,U)\, dP^y(U|E).$$

Using the FDP paradigm for inference, we are able to falsify a subset of assumptions about the unmeasured risk factors based on contrasts of SCC standardized rates. We define two assumptions: **the identical disease probability (IDP)** assumption,

$$(16) \qquad P^{y^\dagger}(D|E,U) = P^{y^{\dagger\dagger}}(D|E,U),$$

and the **comparable-confounding** assumption,

$$(17) \qquad P^{y^\dagger}(U|E) = P^{y^{\dagger\dagger}}(U|E).$$

Without loss of generality, we use as example the inferences that can be made from contrasts of the overall (marginal) SCC standardized cancer rates in years $y^\dagger$ and $y^{\dagger\dagger}$. These are the standardized population cancer rates not conditional on any risk factors ($E_1 = \varnothing$). In terms of the FDP formulation this standardized rate is

$$(18) \qquad s_y^{cc}[D] = \int_E \left\{ \int_U P^y(D|E,U) P^y(U|E) \right\} dP^*(E).$$

If IDP (16) is true, and $s_{y^\dagger}^{cc}[D] \neq s_{y^{\dagger\dagger}}^{cc}[D]$, then we can conclude that the assumption of comparable-confounding (17) is false. For instance, dietary factors such as folate intake are suspected of being risk factors for colon cancer [Giovannucci (2002)]. SEER contains no measurement of folate intake. If within levels of $E$, the intake of folate has changed over time, then (17) is false. In fact, in the United States a population-wide folate supplementation program began in 1998; it is known that folate intake in the population has increased considerably since then [Quinlivan and Gregory (2003)].

Is the IDP assumption reasonable? If we believe that the determinants of a disease, and the impact of those determinants on the probability of disease, are inherent to the biology of humans and do not vary with year, then IDP is true. Such belief is consistent with our current conceptualization of biological processes. However, hidden in the IDP assumption is the assertion that the classification of disease and exposures is identical in year $y^\dagger$ and $y^{\dagger\dagger}$. If diagnostic criteria for colon cancer have changed, or, if diagnostic procedures for detecting colon cancer have changed (for instance, an increase in procedures that lead to early detection of colon cancer), then $D^{y^\dagger}$ and $D^{y^{\dagger\dagger}}$ may not in fact represent the same biological outcome. Similarly, if the measurement tools for ascertaining ethnicity and race have changed, then $E^\dagger$ and $E^{\dagger\dagger}$ may not measure the same attributes. In either case, we would expect IDP (16) to be false.

Identical reasoning proves that if comparable confounding is true (17), then IDP (16) is false.

In summary, if the observed SCC standardized rate differences are nonzero, we can conclude that either IDP (16) and/or comparable confounding (17) are false.

There is a direct correspondence between falsification of the above assumptions and the conclusions made in the *Annual Reports to the Nation*. Ward et al. begin their paper, *Interpreting Cancer Trends*, with the following sentence: "Temporal trends in the incidence of particular types of cancer may reflect changes in exposure to underlying etiologic factors, changes in classification, or the introduction of new screening or diagnostic tests." The "changes in underlying etiologic factors" corresponds to comparable confounding being false (17); the "changes in classification, or the introduction of new screening or diagnostic tests," corresponds to IDP (16) being false.

The FDP inferences given above apply to any SONC operator. They do not apply to the SCA standardization operator used in Ward et al. (2006) or any of the *Annual Reports to the Nation*.

**6. Nested standardized rates and within-year model building.** The *Annual Reports* provide and interpret trends in cancer rates over time for various demographic subgroups. The majority of the report examines contrasts in overall cancer rates, contrasts conditional on gender and race, and contrasts conditional only on gender or only on race. However, unlike in the usual regression analysis, no attempt is made to construct a "parsimonious model." Were the analyses in the *Annual Reports* used only to describe within-group trends over time, such model development might be of no interest. When used to make the type of inferences described in the first paragraph of this paper, the ability to test nested models assumes importance. Whether standardized rates conditional on race and gender are identical to standardized rates conditional on race alone has implications for the allocation of health care resources, the construction of preventive programs, and the focus of future etiologic research.

Regarding standardized rates as the output of standardization operators allows us to evaluate the relationship between standardized rates produced by the same operator on the same data. We define the standardized rate $s_y^*[D|E_1^{\dagger\dagger}]$ to be nested in $s_y^*[D|E_1^{\dagger}]$ provided the following three conditions are true:

(1) both are produced by the same standardization operator, $S_y^*[D|E_1]$,
(2) the arguments of the operator, $(D^y, E^y)$ and $P^*(E)$, are identical,
(3) $E_1^{\dagger\dagger}$ is a proper subset of $E_1^{\dagger}$.

The relationship of nested standardized rates produced by the SCC operator has familiar properties. The SCC operator is recursive in the sense

that

$$(19) \qquad s_y^{cc}[D|E_1^{\dagger\dagger}] = S_y^{cc}[[s_y^{cc}[D|E_1^{\dagger}]]|E_1^{\dagger\dagger}].$$

The right-hand side of (19) is defined to be

$$(20) \qquad S_y^{cc}[[s_y^{cc}[D|E_1^{\dagger}]]|E_1^{\dagger\dagger}] \equiv \int_{E_1^{\dagger} \setminus E_1^{\dagger\dagger}} s_y^{cc}[D|E_1^{\dagger}] \, dP^*(E_1^{\dagger} \setminus E_1^{\dagger\dagger} | E_1^{\dagger\dagger}).$$

The SCC operator does not "discard information." The standardized rate obtained from equation (7) when $E_1 = E_1^{\dagger\dagger}$ is the same estimate obtained by replacing the finest-crude-cancer rate in (7) with $s_y^{cc}[D|E_1^{\dagger}]$. Thus, nested rates produced by the SCC operator have the same properties as nested estimates in regression models of conditional expectations. We are currently developing inferential procedures analogous to those that exist for regression.

Though the identity in (19) can easily be verified by substitution, a more instructive proof is based on the functional form of the SCC operator. We define the probability measure $P^{y*}(D, E) \equiv P^y(D|E)P^*(E)$. The SCC operator can be regarded as the conditional expectation of the finest-crude-cancer rates, $P^y(D|E_1, E_2)$, with respect to $P^{y*}(E_2|E_1)$. Thus, $s_y^{cc}[D|E_1^{\dagger\dagger}]$ is the projection (conditional expectation) of the finest-crude-cancer rates on the subspace (subsigma algebra) defined by $E^{\dagger\dagger}$. Projections (conditional expectations) are entirely determined (a.e. unique) by the subspace (subsigma algebra) on which they are defined (measurable) [Dudley (1989)]. Recursion cannot be defined for SCA. The $P^y(D|E_1^a, A)$ in the integral in (6) is always a function of $A$; $s_y^{ca}[D|E_1]$ is never a function of $A$. Mimicking the form of (20) one might define recursion for SCA to be

$$(21) \qquad \int_{E_1^{\dagger}} s_y^{ca}[D|E_1^{\dagger}] P^*(E_1^{\dagger}|E_1^{\dagger\dagger}).$$

The integral in (21) does not equal the $s_y^{ca}[D|E_1^{\dagger\dagger}]$ obtained from (6). The SCA operator is not a projection operator.

**7. Discussion.** In order to make inference from observational data, researchers attempt to separate the effect of the exposures of interest from the effect of other disease determinants that covary with the exposures of interest. We have divided these other determinants into two mutually exclusive sets: determinants that are measured and determinants that are unmeasured. We refer to procedures that attempt to separate the association of the covariates of interest, $E_1$, from the association of the other measured disease determinants, $E_2$, as procedures that control for confounding.

Standardization is one such procedure. It is the most common procedure used to control for confounding in the analyses of population-studies or data from disease registries. In this paper we examined the ability of various standardization procedures to control for measured risk factors, and the interpretability of differences in standardized rates in the presence of unmeasured risk factors. Our motivation for conducting this research was to evaluate the properties of the "age adjustment using the direct method" standardization procedure (SCA standardization) used in the analysis of SEER cancer registry data and, if needed, to develop standardization procedures with better properties.

We define a general class of standardization operators, and regard standardized rates to be the output from a standardization operator. The general class of operators are any functionals that are integrals of a crude-cancer rate with respect to a user-defined "weighting" distribution (5). Since all crude-cancer rates are themselves integrals of the finest-crude-cancer rates (2), $P^y(D|E)$, standardization operators differ only with respect to the measure used to integrate $P^y(D|E)$. Based on this formulation, we defined between-year differences in crude-cancer rates conditional on $E_1$ as being unconfounded by $E_2$, provided the distribution of $E_2$ conditional on $E_1$ is the same in both years (8). By extension, we defined a subclass of **standardization operators with no $E_2$ confounding** (SONC operators). This subclass consists of standardization operators in which the finest-crude-cancer rates for each $y$ are integrated with respect to a distribution that is the same for all $y$ (A.1).

The SCA operator is not a SONC operator (10). If the differences in crude-cancer rates are not confounded, the SCA operator can introduce confounding and produce between-year differences in standardized rates that are confounded. In Section 3.1 we showed that the SCA operator will introduce confounding unless the $P^y(E)$ distributions differ only in terms of the marginal distribution of age. This criterion was not met in the SEER 13 data that we analyzed. Figure 1 from that analysis provides a graphic representation of the fact that the standardized rates produced by the SCA operator for year $y$ are not the "cancer rates one would have seen" had the age distribution in year $y$ been identical to the age distribution that generated the weights.

It is clear that one should always choose a standardization operator from the subclass of operators with no $E_2$ confounding. We proposed and examined the properties and performance of one such operator: the standardization controlling for covariates operator (SCC). A desirable characteristic of the SCC operator is that the year-to-year differences in standardized cancer rates preserve the magnitude of the differences of the crude-cancer rates. When the crude-cancer rates are not confounded, and thus no standardization procedure is required, the SCC operator is the only operator for which the standardized rates equal the crude-cancer rates. Figure 1 provides a graphic illustration of these properties. Figure 2 shows that the largest differences between standardized rates from the SCA operator and crude-cancer rates occurs in minority populations.

If standardized rates produced by standardize operators with no $E_2$ confounding are different in year $y^\dagger$ and $y^{\dagger\dagger}$, then differences must exist in the finest-crude-cancer rates. However, to make the inferences of interest to the authors of the *Annual Reports* [Ward et al. (2006)], one must consider the impact of unmeasured cancer risk factors on the year-to-year differences in standardized rates. In Section 5 we used the fundamental disease probability paradigm for inference proposed by Mark (2004, 2005, 2006, 2008) to prove that nonzero contrasts of standardized rates falsify assumptions about the conditional probabilities of disease given all the risk factors [the **identical disease probability assumption**, (16)], and/or the conditional probabilities of unmeasured risk factors given the measured risk factors [the **comparable-confounding assumption**, (17)]. We describe the one-to-one correspondence that exists between the inferences made in the SEER *Annual Reports* [Ward et al. (2006)], and the falsification of these assumptions.

The analyses in the *Annual Reports* only examine between-year differences in cancer rates. In Section 6 we argue that, given the type of inferences made from these reports, it would be desirable to examine within-year contrasts. We defined the concept of nested standardized rates. We proved that, like regression models for conditional expectations, the SCC operator is a projection operator. In our current research we are developing methods for testing nested models analogous to those used in regression.

Though we have discussed the SCC operator in terms of estimating the conditional expectation of a binary variable, these operators extend in an obvious manner to the estimation of conditional expectations in general. The particular operators we have defined are nonparametric estimators. One could construct parametric or semiparametric operators by, for instance, replacing $P^y(D|E)$ with a parametric or semiparametric model, $P^y(D|E; \theta)$.

The next step in our research is to derive SCC estimators for the statistics currently used for inference in the *Annual Report*. Our eventual goal is to program theseestimators in the $R$-language [R Development Core Team (2007)] and produce freely available user-friendly software comparable to the SEER*Stat freeware [SEER*Stat.6.3.6. (2007)] available for SCA analyses of SEER data.

## APPENDIX: DEFINING STANDARDIZATION OPERATORS WITH NO $E_2$ CONFOUNDING

For all factorizations $E = (E_1, E_2)$, let $F^*(E_1, E_2)$ be a family of probability measures indexed by $e_1^*$, with the property that $\int_{\mathcal{E}_2} dF^*(e_1^*, E_2) = 1$ for all $e_1^* \in \mathcal{E}_1$.

$S_y^*[D|E_1]$ is a standardization operator unconfounded by $E_2$ if it can be expressed in the form

$$S_y^*[D|E_1] = \int_{\mathcal{E}_2} P^y(D|E_1, E_2) \, dF^*(E_1, E_2). \tag{A.1}$$

Note that for the SCC operator (7), the $P^*(E)$ specifies the $F^*(E_1, E_2)$ for all factorization of $E$, and all realizations $e^* \in \mathcal{E}$. In general, this need not be the case.

## SUPPLEMENTARY MATERIAL

**Fundamental disease probability inference: A new paradigm for causal inference in the biological sciences** (DOI: 10.1214/08-AOAS170SUPP; .pdf).

Department of Biostatistics and Biometrics
University of Colorado
School of Public Health
4200 East Ninth Avenue, RM1615
Denver, Colorado 80262
USA
E-mail: steven.mark@uchsc.edu